%% file: nooyens2025ictss.tex
\documentclass[runningheads]{llncs}
\input{usecommands}
\input{usepackages}
\input{usesettings}
\usepackage[T1]{fontenc}
\usepackage{graphicx}
\begin{document}
\title{Test Amplification for REST APIs via Single and Multi-Agent
LLM Systems\thanks{University of Antwerp and Flanders Make}}
%
%
\author{Robbe Nooyens\inst{1}\orcidID{0009-0003-4800-1651} \and
Tolgahan Bardakci\inst{1}\orcidID{0009-0007-1136-2065} \and
Mutlu Beyaz{\i}t\inst{1}\orcidID{0000-0003-2714-8155} \and
Serge Demeyer\inst{1}\orcidID{0000-0002-4463-2945}}
\authorrunning{R. Nooyens et al.}
%
\institute{University of Antwerp 2020, Antwerp Belgium
\email{robbe.nooyens@student.uantwerpen.be}
\email{tolgahan.bardakci@uantwerpen.be}
\email{mutlu.beyazit@uantwerpen.be}
\email{serge.demeyer@uantwerpen.be}\\
}
\maketitle              
\begin{abstract}
REST APIs (Representational State Transfer Application Programming Interfaces) play a vital role in modern cloud-native applications.
As these APIs grow in complexity and scale, ensuring their correctness and robustness becomes increasingly important. Automated testing is essential for identifying hidden bugs, particularly those that appear in edge cases or under unexpected inputs.
However, creating comprehensive and effective test suites for REST APIs is challenging and often demands significant effort.
In this paper, we investigate the use of large language model (LLM) systems—both single-agent and multi-agent setups—for amplifying existing REST API test suites.
These systems generate additional test cases that aim to push the boundaries of the API, uncovering behaviors that might otherwise go untested. We present a comparative evaluation of the two approaches across several dimensions, including test coverage, bug detection effectiveness, and practical considerations such as computational cost and energy usage.
Our evaluation demonstrates increased API coverage, identification of numerous bugs in the API under test, and insights into the computational cost and energy consumption of both approaches.

\end{abstract}
\section{Introduction}
REST APIs (Representational State Transfer Application Programming Interfaces) are crucial for today's cloud applications.
They offer high performance and flexibility, key components for efficient, scalable cloud services.
Although their benefits are undeniable, they come with certain challenges~\cite{verborgh2015fallacy}.

REST APIs are inherently distributed, making it challenging to ensure their overall quality. 
The vast number of possible interactions among protocols and the even greater variety of API call sequences imply a sheer infinite amount of possible input, where only a few will expose a bug, the famous ``needle in a haystack'' phenomenon. 
The testing challenge is further increased by the involvement of multiple engineering teams developing different components independently; as such, parts of the protocol evolve at different speeds.
Therefore, testing the different aspects, especially performing boundary value testing, is essential as it is the best possible answer to expose lurking defects.

\paragraph{Boundary value testing} is a software testing technique that focuses on inputs at the edge of valid input ranges where systems are most likely to fail.
Guo et al.~\cite{GuoOkamuraDohi2024} introduce a metric called boundary coverage distance (BCD) to quantify how well a test suite covers these critical boundary points. 
Their experiments show that test suites optimized for BCD are more effective at detecting faults than traditional random testing methods.

\paragraph{Test amplification} is a likely solution for searching the needle in the haystack, as substantial evidence supports its effectiveness in unit tests~\cite{test_amplification_definition}.
While test generation is also useful, its lack of test suite context can lead to redundant or irrelevant tests.
Test amplification, by contrast, retains contextual relevance by reusing and refining existing test logic.
It aligns more closely with how developers write and maintain tests in practice.
This aspect of test amplification makes it especially appealing in real-world projects, where developer trust and adoption are key.
Studies show that test amplification enhances the test suite based on engineering goals, including, but not limited to, increasing coverage, reducing the impact of software defects, and improving mutation analysis~\cite{test_amplification_ampyfier}~\cite{schäfer2024empiricalevaluationusinglarge}.

\paragraph{The use of artificial intelligence,} especially large language models (LLMs), is growing rapidly in software engineering~\cite{ahmed2023survey}~\cite{vaithilingam2022prompts}~\cite{zhang2023chatgpt}.

Multiple studies showed the potential of LLMs for unit test amplification.
For example, a study carried out within META shows promising results in improving unit tests using LLMs~\cite{unit_test_amplification-meta}.

Bardakci et al. demonstrated the potential of test amplification with LLMs, where out-of-the-box LLMs are used to perform test amplification for REST APIs~\cite{bardakci2025testamplificationrestapis}.
In this paper, we continue this research by exploring the feasibility of using agentic LLM systems for the same task.
The motivation for this is twofold.
First, recent work in single-agent REST API test generation has explicitly stated the use of multi-agent systems as a promising research direction for future work~\cite{pereira2024apitestgenieautomatedapitest}.
Second, the idea of decomposing complex software engineering challenges into specialized roles for collaborating agents has proven highly effective in areas such as software development~\cite{hong2024metagptmetaprogrammingmultiagent} and automated test generation~\cite{stennett2025autoresttesttoolautomatedrest}.
The study of AgentCoder~\cite{huang2024agentcodermultiagentbasedcodegeneration} showed that splitting the programmer and test designer into multiple agents achieved a 13\% to 15\% increase in coverage compared to a single agent.
However, given the added complexity and overhead, it remains unknown whether a multi-agent system is suitable for a test amplification task.
Therefore, this paper presents an initial feasibility study to explore the characteristics, potential, and trade-offs of the single- and multi-agent architectures in this specific context.

\section{Related Work}

Test amplification has proven to be a successful approach as an advanced testing method.
There is substantial evidence suggesting that this approach is highly effective and multiple studies have demonstrated its reliability and value across various contexts~\cite{test_amplification_definition}~\cite{test_amplification_ampyfier}~\cite{schäfer2024empiricalevaluationusinglarge}.

Leveraging LLMs for software testing purposes shows promising results.
For instance, TestGen-LLM~\cite{unit_test_amplification-meta} leverages LLMs for unit test amplification.
The researchers demonstrate that the TestGen-LLM has 75\% build accuracy with 57\% passed reliably and 25\% increased coverage.
Yang et al.~\cite{yang2024evaluationlargelanguagemodels} used open-source LLMs to generate unit tests.
They found that the way the prompts are structured is important and noticed that giving the LLM less information can sometimes work better, which suggests that clear and simple task instructions might be more effective.
Similarly, Schäfer et al.~\cite{schäfer2024empiricalevaluationusinglarge} utilized LLMs to generate unit tests automatically and added a way to fix errors by feeding the model the failing tests and error messages to improve the tests.
Yang et al.~\cite{yang2024enhancingllmbasedtestgeneration} introduced TELPA, a method that uses program analysis and counterexamples in prompts to help LLMs understand the target method better and iteratively refine counterexamples to reach hard-to-cover branches.
Pizzorno and Berger~\cite{pizzorno2025coverupcoverageguidedllmbasedtest} proposed CoverUp, an LLM-based test generator that significantly improves Python code coverage by combining code context, coverage analysis, and iterative feedback.

Agentic LLM frameworks have been developed and proven effective in REST API testing.
APITestGenie~\cite{pereira2024apitestgenieautomatedapitest} is a single-agent tool that generates REST API test cases and incorporates local execution feedback to validate the generated tests.
AutoRestTest~\cite{stennett2025autoresttesttoolautomatedrest} successfully generates request headers, parameters, and parameter values with specialized agents to detect unexpected behavior and bugs.
The results demonstrate the strength of LLMs and agentic systems in improving the robustness of software testing.
In addition, regardless of single or multi-agent settings, they offer an opportunity to improve the quality of the systems significantly.
Furthermore, they provide valuable ideas that can be extended to test amplification.

An empirical study by Kim et al.~\cite{Kim_2022} provides a comprehensive overview of automated REST API testing techniques.
These include evolutionary-based testing, which uses evolutionary algorithms to generate and refine effective test cases.
Dependency-based algorithms infer relationships between operations to produce stateful tests.
Model-based testing uses abstract models of the API that define constraints among parameters to generate test cases.
Property-based testing defines invariants to verify API behavior.
Random testing uses random assignments for input parameters.
Combining this with mutation techniques is known as random-mutation-based testing.
Sample value testing relies on example or default values in the OpenAPI documentation.
Robustness testing tries to find issues by generating valid and invalid inputs.

\hypobox{
To the best of our knowledge, there are no agentic LLM systems designed for REST API test amplification.
We demonstrate a novel approach that leverages single and multi-agent systems to increase API coverage.
Our outcomes show a significant increase in API coverage and many bug identifications in the targeted cloud application.}

\section{Approach}
This section provides fundamental background information to formulate our approaches and then introduces our agentic LLM systems for test amplification.

\subsection{Background}

\paragraph{REST APIs} are a web-based interfaces that allow clients to access and manipulate resources on a server using standard HTTP methods.
The REST principles emphasize stateless communication, resource-oriented architecture, and a uniform interface using URLs to represent data entities~\cite{fielding2000rest}.

\paragraph{Test amplification} is an umbrella term for various activities that analyze and operate on existing test suites, including augmentation, optimization, enrichment, and refactoring~\cite{test_amplification_definition}.
Test amplification is distinct from test generation.
It produces additional test cases derived from existing ones rather than constructing them from scratch.
This approach provides a significant advantage since the amplified tests align more closely with the established test architecture.

\paragraph{Large Language Models (LLMs)} are advanced artificial intelligence models trained on vast amounts of text data to understand, generate, and manipulate human language.
These models, typically based on deep learning architectures, can perform a wide range of language-related tasks, such as code generation.

\paragraph{Generative Pre-trained Transformer (GPT)} is an advanced LLM developed by OpenAI, based on the GPT architecture~\cite{zhao2025surveylargelanguagemodels}.
It is designed to understand and generate human-like text, enabling natural conversations and assisting various language-related tasks.
A wide range of applications and research proves GPT's capability in the software engineering world~\cite{schäfer2024empiricalevaluationusinglarge}.

\paragraph{Agentic LLM systems} are designed around the idea of autonomous agents.
They have components powered by language models that can make decisions, take actions, and adapt based on their given tasks.
They operate independently, reasoning through problems, interacting with tools or environments, and producing meaningful outputs based on their goals and inputs.
These systems fall into two types: single and multi-agent systems.
In a single-agent system, one agent handles the entire set of tasks on its own.
In multi-agent systems (MAS), a team of specialized agents collaborates, each contributing their strengths towards a shared objective~\cite{bousetouane2025agenticsystemsguidetransforming}.

\subsection{Baseline: Single-Agent Setup}
Pereira et al. (2024) developed a tool called APITestGenie~\cite{pereira2024apitestgenieautomatedapitest}.
They show that single-agent LLM architectures can be leveraged for generating executable REST API test scripts with minimal human intervention, achieving a success rate of up to 80\%.
Since APITestGenie is designed for test generation and it has different requirements to operate, we built our own test amplification system with a single-agent, similar to how APITestGenie works.
This will be our baseline due to its simplicity, ease of implementation, and effectiveness, as evidenced in prior work.

\begin{figure}[ht]
    \centering
    \includegraphics[width=0.4\textwidth]{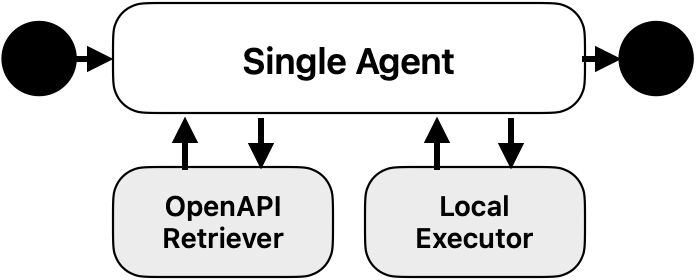}
    \caption{Single-Agent Setup}
    \label{fig:single-agent}
\end{figure}

Both of our systems are implemented using the LangGraph library~\cite{LangGraph}.
We chose LangGraph as it extends the popular open source LangChain framework and it is specifically designed for building stateful, agentic applications with explicit control over the workflow.
For the single-agent workflow, we use LangGraph's built-in ReAct (Reasoning and Acting) agent.
This agent architecture is designed to reason about a task and decide when to use external tools to retrieve information or perform actions~\cite{yao2023reactsynergizingreasoningacting}.
The agent has access to two specialized tools, as shown in \autoref{fig:single-agent}.
The ``OpenAPI Retriever'' is a tool that can be used to query information from the OpenAPI documentation for a given endpoint or definition.
The agent can query details about endpoints (identified by a '/'), such as available operations, parameters, request/response types, and defined status codes.
It can also retrieve information about specific data definitions, such as the structure and properties of objects like 'Order', 'User' or 'Pet'.
This allows the agent to understand the API's requirements, constraints and expected behavior.
The ``Local Executor'' saves the generated tests in a testing repository and runs them in a local environment.
The execution output from the console reveals information about passing and failing tests, as well as syntax and runtime errors.

The effectiveness of such an agent is highly dependent on its instruction.
The importance of prompt design is well-documented, with studies showing that prompt structure significantly impacts performance~\cite{yang2024evaluationlargelanguagemodels}.
They found that preliminary experiments are necessary for determining an optimal prompt and that removing useless information can be beneficial.
Based on this principle, our prompt was engineered to provide extensive but concise context, including details about the testing framework, available functions, and local files.
It explicitly instructs the agent to maximize REST API coverage across multiple criteria~\cite{test_coverage_criteria} and to explore edge cases to find bugs.
The complete prompt, which was iteratively refined to be both extensive and effective, can be found in the web appendix, "section A"~\cite{nooyens2025test_appendix}.

\subsection{Multi-Agent Setup}

While a single-agent system provides a strong baseline, the literature on agentic systems suggests that complex tasks often benefit from breaking down a problem and assigning it to specialized agents~\cite{hong2024metagptmetaprogrammingmultiagent}.
These agents, with their own clear and concise instructions, can efficiently handle distinct responsibilities in the test generation workflow.
By giving each agent a focused prompt for a smaller sub-task, we hypothesize that the overall system can perform better than a single agent handling a complex, multi-part prompt~\cite{yang2024evaluationlargelanguagemodels}.

\begin{figure*}[htbp]
    \centering
    \includegraphics[width=0.9\textwidth]{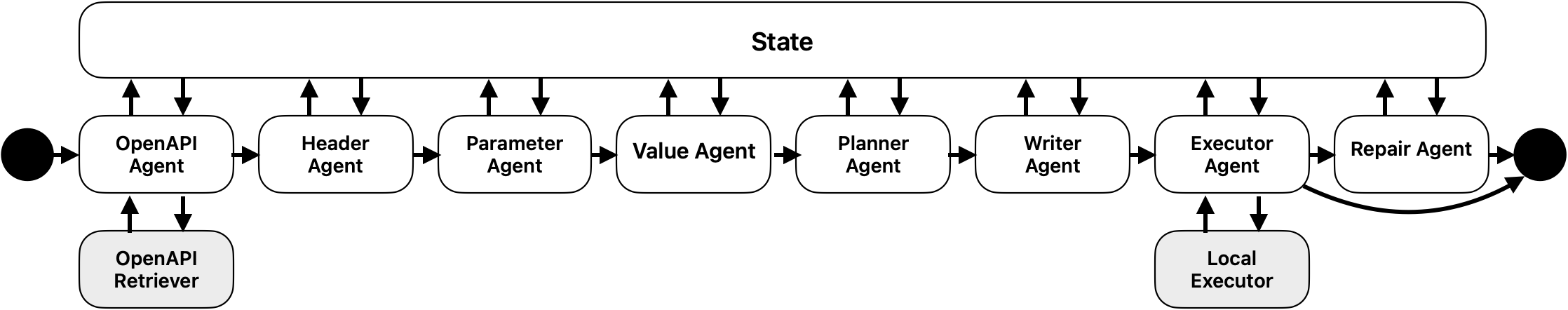}
    \caption{Multi-Agent Setup}
    \label{fig:multi-agent-chain}
\end{figure*}

Our multi-agent architecture, shown in \autoref{fig:multi-agent-chain}, was designed by recursively decomposing the task of "amplifying a test suite".
This led to a sequential pipeline with two phases, being test case planning and test code generation.
This separation of concerns (first designing what to test, then writing the code), is a proven principle in multi-agent software engineering systems that separate high-level design from code implementation~\cite{hong2024metagptmetaprogrammingmultiagent}~\cite{huang2024agentcodermultiagentbasedcodegeneration}.

\paragraph{The planning phase} is further decomposed into specialized roles.
To ensure the planning agents have the necessary context, the first agent is the OpenAPI Agent, who is responsible for retrieving relevant sections from the OpenAPI specification.
This context is passed to a set of agents inspired by the approach in AutoRestTest~\cite{stennett2025autoresttesttoolautomatedrest}: the Header Agent, Parameter Agent and Value Agent.
These are prompted to generate suggestions for inputs that target specific REST API coverage criteria~\cite{test_coverage_criteria}, such as values that trigger documented status codes or boundary values for parameters.
Finally, the Planner Agent combines the suggestions from these parallel agents into a list of test case descriptions.
For large and detailed OpenAPI specifications, the context we provide for the agents can be quite long. 
Since GPT updates its hidden state from left-to-right~\cite{zhao2021calibrateuseimprovingfewshot}, we provide the entire context before the instructions, so the model can build an internal representation based on the specific context as it processes the input sequentially.
We also instructed the agents to use Chain-of-Thought reasoning, a prompting technique that encourages the model to break down a problem into a set of intermediate steps before providing a final answer~\cite{wei2023chainofthoughtpromptingelicitsreasoning}.
This is useful as it can help them to infer non-obvious combinations of parameters and values that could trigger specific API behaviors.
The exact input prompts can be found in the web appendix, "section A"~\cite{nooyens2025test_appendix}.

\paragraph{The generation phase} takes the plan and turns it into executable code.
This phase is inspired by feedback-driven frameworks, like CodeCoR~\cite{pan2025codecorllmbasedselfreflectivemultiagent}, which uses a Test Repair Agent to propose code changes based on the output from a local execution.
We adapt this principle by using a Test Writer Agent to translate the test descriptions from the Planner into a code file.
The Test Executor Agent validates the code by running it.
If the agent detects compilation or runtime errors, it passes the error feedback to a dedicated Test Repair Agent, who is prompted to fix the original code.
Their input prompts can be found in the web appendix, "section A"~\cite{nooyens2025test_appendix}.

\paragraph{The workflow} defines the sequence of agent interactions and the conditions for transitioning between agents and tools, also orchestrated by the LangGraph library~\cite{LangGraph}.
For example, if the OpenAPI Agent needs to use the OpenAPI retriever tool, the workflow transitions to a tool node, which executes the tool and returns the results to the OpenAPI Agent.
Similarly, the workflow includes a conditional path to the Test Repair Agent, which is triggered only if the Test Executor Agent detects errors that need to be fixed.
This structured multi-agent approach results in a more systematic amplification of REST API tests, with each agent contributing its expertise to different aspects of the testing creation process.
Furthermore, the workflow operates as a pipeline without a feedback loop from the execution results back to the planning phase.
This ensures deterministic evaluation and manages the complexity of the initial feasibility study.
Integrating an automated coverage analysis tool that automatically captured coverage reports from Restats~\cite{restats} was not possible with our experimental setup, which required manual steps to obtain those.
However, the modular design implemented in LangGraph allows for easy addition of such loops in future work, where a coverage report from the Test Executor could be fed back to the Planner to guide the generation of subsequent tests.

\section{Evaluation Setup}
Our evaluation aims to validate the feasibility of utilizing agentic large language model (LLM) systems for REST API test amplification. To achieve this, we aim to address the following research question.

\begin{itemize}
\item \textit{How do single-agentic and multi-agentic LLM systems perform at REST API test amplification comparatively?}
\end{itemize}

\subsection{System Under Test}
We validate our approach using the PetStore~\cite{SwaggerPetstore} application.
PetStore is a widely recognized representative cloud application with several suitable characteristics for our study.
It is an open-source system that exposes multiple API endpoints, supporting essential operations such as reading, writing, updating, and deleting data.
It contains twenty different endpoints and supports various parameters for interaction with those endpoints.
It also includes both JSON and XML request and response types.
Considering those traits, it introduces complexity into the API structure, making the structural API coverage metrics more meaningful.

Furthermore, the Petstore API presents common and important API testing challenges, such as parameter constraints, file uploads, and data dependencies, all of which are highly relevant to real-world API testing scenarios.
The application allows us to evaluate the effectiveness and reliability of our approaches in a REST API environment.

\subsection{Approach Parameters}
We use the following parameters in all of our approaches.
\begin{itemize}
    \item We choose OpenAI's GPT-4o mini model as our LLM due to its optimal balance of quality and competitive pricing.
    \item A temperature closer to 0 makes the model more deterministic, while a value closer to 1 increases creativity and variability in its responses.
    We set the temperature parameter to zero to reduce randomness and ensure more predictable outputs that closely follow our instructions and the OpenAPI documentation.
    \item We provide an initial happy-path test case which is considered as our test suite, along with function definitions, for the testing framework's GET, POST, DELETE, and PUT methods.
    \item We create essential image files to test specific endpoints and supply them to our systems.
    LLMs utilize these image files to enhance accuracy in data usage during the amplification process.
    \item We run our systems three times to account for variations in LLM outputs and get the average values of our runs.
\end{itemize}

All materials, including parameters, are available in our reproduction package\footnote{\url{https://figshare.com/s/15158c0adf43b5410188}}.

\subsection{Evaluation Criteria}
To answer our research question, we evaluated the results in four distinct categories of metrics, inspired by established practices in REST API testing literature.
We chose test execution statistics to measure direct outputs and fault detection capabilities of our systems ~\cite{pereira2024apitestgenieautomatedapitest}, structural API coverage to quantify how much has been covered from a black-box perspective ~\cite{restats} \cite{Kim_2022}, LLM usage statistics to assess the practical cost and efficiency of employing generative models ~\cite{pereira2024apitestgenieautomatedapitest}, and readability to evaluate the qualitative aspects and maintainability of generated code~\cite{buse2010learningametricforcodereadability}, a key strength of Large Language Models ~\cite{schäfer2024empiricalevaluationusinglarge}.

\paragraph{Test execution statistics} cover the fundamental observations made using the amplified test cases.
We report the number of tests generated, along with counts of successful, failed, and bug-exposing tests.
The validity and categorization of test cases are determined by a manual review process which was conducted by one of the authors and verified by another.
Additionally, each generated and failed test case was individually assessed and assigned one of the following labels.
Those labels are based on the APITestGenie~\cite{pereira2024apitestgenieautomatedapitest}.
\begin{itemize}
    \item API Bug Exposed: This label was assigned when a test case revealed an inconsistency between the API's documented behavior (according to the OpenAPI documentation) and its actual behavior.
    This includes cases where the server responded with an incorrect error code or returned a success code where an error was expected.
    \item Missing Information: Test cases that made assumptions about the existence or non-existence of specific data, such as assuming a particular ID would or would not exist, when that information is not specified in the OpenAPI documentation.
    For example, a test might assume that ID 9999 is invalid, but the API might accept it.
    \item Semantically Incorrect: This new category contains test cases that check assertions not documented in the OpenAPI specification.
    These tests failed because they were checking for something the API was not designed to do.
    \item Runtime/Syntactical Error: Test cases that caused runtime exceptions or compilation errors.
\end{itemize}

\paragraph{Structural API coverage} values are calculated for path, operation, status class, status, response type, request type, and parameter coverage~\cite{test_coverage_criteria}.
We use the Restats~\cite{restats} tool to calculate the REST API coverage automatically.

\paragraph{LLM usage statistics} includes the token usage, runtime, cost of usage and energy consumption.
The cost of usage can be estimated by multiplying the amount of input and output tokens by OpenAI's pricing rate.
We obtained these values through LangSmith~\cite{langsmith2023}, an open source application that gathers these statistics.
Metrics like token usage, runtime and cost have been evaluated in prior work before~\cite{pereira2024apitestgenieautomatedapitest}.
In addition, we evaluate energy consumption.
With the growing computational scale of AI, understanding the energy cost becomes more and more important for evaluating the sustainability of such approaches.
The energy consumption is calculated as follows:
A typical GPT-4o query consumes approximately 0.3 watt-hours of energy for about 500 tokens~\cite{epoch2024chatgptenergy}.
GPT-4o-mini, with around 8 billion parameters versus GPT-4o’s 200 billion, is estimated to consume less energy, about 0.03 watt-hours for a 500-token query, or roughly 0.00006 watt-hours per token.

\paragraph{Readability} is an important non-functional requirement in software engineering that influences maintainability~\cite{buse2010learningametricforcodereadability}.
To evaluate readability, we assessed the generated test cases against a checklist.
The checklist consists of the following criteria:
\begin{itemize}
    \item Meaningful Naming: Are the generated test method and variable names descriptive and consistent with the conventions of the target programming language?
    \item Structural Coherence: Is the generated test case consistent with the style of the initial example test?
    \item Idiomatic Correctness: Does the code use standard features of the target language and testing framework effectively?
\end{itemize}

The readability assessment process begins with a review conducted by an author. 
Another author is assigned to review the readability after the first author completes their work.
Due to the qualitative aspect of readability, it is best to share examples with readers.
These examples can be accessed from our web appendix, "section B"~\cite{nooyens2025test_appendix}.

\section{Results and Discussion}
\paragraph{Our test execution statistics} are given in \autoref{tab:test_statistics}.
The multi-agent approach yields a greater number of test cases overall.
Although it mostly contributes to the failed test cases, the majority of these failed test cases are not invalid; they expose bugs that need to be addressed.

For example, consider the generated test case in the web appendix "section B.2", which is a bug-exposing test case.
This test case is designed to verify the API's response when a user attempts to log in with an invalid password.
According to the OpenAPI specification, the call should return 400 when "Invalid username/password supplied".
However, a success status code was returned, exposing an inconsistency between the OpenAPI specification and the actual behavior.

\begin{table}
\centering
\caption{Test Execution Statistics}
\label{tab:test_statistics}
\begin{tabular}{|l|c|c|}
\hline &               {\bfseries Single-Agent} & {\bfseries Multi-Agent}\\ \hline
Generated Tests        & 118                    & 145                     \\
Successful Tests       & 58 (49\%)              & 67 (46\%)               \\
Failed Tests           & 60 (51\%)              & 78 (54\%)               \\
Bug-exposing Tests     & 32 (27\%)              & 43 (30\%)               \\
\hline
\end{tabular}
\end{table}



\autoref{tab:failedtestcategories} categorizes the failed tests based on the manual review process mainly into two categories, valid and invalid, where invalid test cases are further divided into three categories. All these categories are mutually exclusive.

An example of a valid bug-exposing test case can be found in the web appendix "section B.2", a failing test case due to missing information in "section B.3, Listing 12", a semantically incorrect test case in "section B.3, Listing 13" and a test case that throws a runtime error in "section B.3, Listing 14"~\cite{nooyens2025test_appendix}.
\autoref{tab:failedtestcategories} further shows that the multi-agent approach generates a greater number of tests that expose bugs, which have missing information, and which are semantically incorrect. It generates fewer tests that have compilation or runtime errors, indicating a higher quality of generated test code.

\begin{table}
\centering
\caption{Categorization of Failed Tests}
\label{tab:failedtestcategories}
\begin{tabular}{|l|c|c|}
\hline                                  & {\bfseries Single-Agent} & {\bfseries Multi-Agent} \\ \hline
Failed Tests                            & 60                       & 78                      \\
Valid - Bugs exposed                    & 32 (53\%)                & 43 (55\%)               \\
Invalid - Missing Information           & 16 (27\%)                & 23 (30\%)               \\
Invalid - Semantically Incorrect        & 4  (7\%)                 & 8  (10\%)               \\
Invalid - Compilation or Runtime Errors & 8  (13\%)                & 4  (5\%)                \\
\hline
\end{tabular}
\end{table}

\paragraph{Structural API coverage} achieved by different setups are shown in \autoref{tab:structural_API_coverage}.
Initial coverage values belong to our test case, which was written by a developer implementing a happy-path scenario  t(web appendix "section B.1, Listing 10")~\cite{nooyens2025test_appendix}.
After leveraging the example test case in our systems (in a fully automated manner), both systems increased the coverage significantly.
Although our more advanced multi-agent system performs better in most cases, the single-agent system still shows promising results.

In our system, we have twenty different paths, as indicated by endpoint numbers.
During the runs, we loop over each endpoint; thus, path coverage easily hits 100\%.
The real difference lies in areas such as Response Type, Request Type, and Parameter coverage.
The multi-agent system outperforms the single-agent system in those areas.
The specialized agents in the multi-agent setup can explicitly choose a targeted set of values to test, resulting in a more specific task description for the planner and increased coverage.

Success status codes that require a valid input ID may not be triggered during testing because the LLM must guess which IDs exist in the database.
Conversely, some tests that were expected to result in a bad request instead returned success codes.
These cases reveal potential bugs in the implementation.
As the test coverage improves significantly, we observe that some status codes are triggered even though they are not documented in the OpenAPI specification.
This indicates that the documentation may be incomplete or outdated.
These uncovered cases highlight the additional benefits of comprehensive software testing.

\begin{table}
\centering
\caption{Structural API Coverage}
\label{tab:structural_API_coverage}
\begin{tabular}{|l|c|c|c|c|}
\hline                  & {\bfseries Initial Coverage}        & {\bfseries Single-Agent}       & {\bfseries Multi-Agent}    \\\hline 
Path             &7\%                       & 100\%              & 100\%           \\
Operation        &5\%                       & 95\%               & 100\%           \\
Status Class     &4\%                       & 48\%               & 54\%             \\
Status           &3\%                       & 36\%               & 37\%             \\
Response Type    &3\%                       & 51\%               & 88\%              \\
Request Type     &9\%                       & 73\%               & 97\%              \\
Parameter        &11\%                      & 56\%               & 82\%               \\
\hline
\end{tabular}
\end{table}

\paragraph{LLM usage statistics} are outlined in \autoref{tab:llmstatistics}.
These statistics provide insights into several key dimensions: time, token usage, cost, and energy consumption. 
In terms of financial cost, generating $118$ test cases using a single-agent setup costs $\sim0.02$ dollars, while generating $145$ test cases with a multi-agent system costs $\sim0.05$ dollars.
Although the number of generated tests is slightly higher for the multi-agent approach, the relative cost increase is about $2.8$ times more expensive than the single-agent method.

Energy consumption also shows a similar trend. The multi-agent approach consumes roughly $2.3$ times more energy than the single-agent system.
This increase is primarily due to the collaborative nature of the multi-agent system.
The multi-agent system requires more computation, coordination overhead, and reasoning steps among agents. Additionally, the increased token usage in multi-agent setups directly contributes to both higher costs and greater energy consumption.

These results show that while multi-agent systems consume more time, tokens, cost, and energy, they may be justified when higher test quality or coverage is required.
The added cost can be worthwhile for more robust outcomes, but it’s important to weigh these benefits against available resources and performance needs.

\begin{table}
\centering
\caption{LLM Usage Statistics}
\label{tab:llmstatistics}
\begin{tabular}{|c|c|c|}
\hline  & {\bfseries Single-Agent}     & {\bfseries Multi-Agent}               \\ \hline
Time (s)                               & 351                       & 1149       \\
Tokens                                 & 71186                     & 162355     \\
Cost of Usage (\$)                     & 0.01848                   & 0.05213    \\
Energy Usage (Wh)                      & 4.27                      & 9.74       \\
\hline
\end{tabular}
\end{table}

\paragraph{Readability} is one of the strong suits of LLMs in programming. 
They are good at producing natural-looking completions for source code, and to some extent understand the semantics~\cite{schäfer2024empiricalevaluationusinglarge}.
Applying our readability checklist from section 4.3, we found that the generated code was consistently high quality.
We present an analysis of two amplified test cases.

The first amplified test case can be found in the web appendix "section B.4, Listing 15"~\cite{nooyens2025test_appendix}.
We have a concise test scenario created by the LLMs that pushes the system under test to its limits.
In this case, the LLM tests the system by leaving out a required parameter.
The provided happy path test case (web appendix "section B.1, Listing 10")~\cite{nooyens2025test_appendix} uses the Arrange-Act-Assert pattern and defines a separate map for header values.
This design is clearly reflected in the generated test case.
Additionally, the function name in Java's camel case describes the functionality being tested quite well, a new variable is created with a proper name for the petId argument, and it correctly uses the "@Test" annotation and assert statement.
It also includes a comment that encourages developer involvement and suggests potential improvements to the test case.

Another example of an amplified test case is shown in the web appendix "section B.4, Listing 16"~\cite{nooyens2025test_appendix}.
This test is well-constructed for several reasons.
First, it builds valid JSON and XML payloads that match the API's expected format.
This demonstrates the models’ understanding of content types.
Second, it uses the ``DataProvider'' annotation to iterate over these types, without causing compilation errors.
Much like the previous example, the Arrange-Act-Assert pattern is present and the name of this test case also clearly describes what will be tested.
To differentiate between the two responses, it successfully added a prefix to both response variable names.
The test also uses the correct data structure and includes clear assertions to verify both status codes and the content of the response body.

Both systems perform well with respect to the readability criteria we defined earlier.
The meaningful naming, structural coherence, and idiomatic correctness were consistent across the majority of the generated tests we reviewed.
We can conclude that both single-agent and multi-agent systems produce well-designed and maintainable test cases, delivering improvements not only quantitatively but also qualitatively.

\section{Threats to Validity}
Several factors may impact the outcome of our approach and, consequently, the validity of our research.
We categorize these threats into four categories.





\subsection{Construct Validity}

We use structural API coverage and the number of bugs found as primary indicators for the "effectiveness" of test amplification.
While these are standard metrics in the field~\cite{Kim_2022}, they have limitations. High API coverage does not guarantee that the most critical or complex parts of the API have been tested, only that they have been reached.
Similarly, our bug count treats all identified issues as equal, without distinguishing between critical and minor bugs.

\subsection{Internal Validity}

The primary internal threat is the LLM's potential prior knowledge of the system under test.
Since PetStore is a widely used, open-source application, the model may have been trained on data related to it, including its documentation or even existing test cases.
To mitigate this concern, we have designed an automated system that functions across multiple cloud applications.
The goal of this work is an initial feasibility study, and using a well-known application like PetStore is appropriate for this stage.
An additional mitigation strategy would involve using local LLMs or alternative cloud-based models that have neither prior nor current access to the source or test code of the system under test.

\subsection{External Validity}

Our evaluation is conducted on a single application, PetStore.
While it is a representative system with a variety of endpoints and operations, the results may not directly generalize to other APIs with different characteristics, such as more complex authentication schemes or different architectural styles.
To address this point, we developed a fully automated approach that can be easily applied to other cloud applications.
To further enhance the general adaptability of this approach, future work may include replacing manual code review with automated evaluation tools, potentially powered by LLMs or other approaches.

\subsection{Reliability}

LLms are inherently non-deterministic and can produce different outputs on identical inputs.
Therefore, identical reproduction of the experiments is not guaranteed across different runs.
To mitigate this, we ran each experiment three times and reported the average values to ensure the results are representative of expected performance.

Our results are obtained using the OpenAI GPT-4o mini model (with around 8 billion parameters compared to the estimated 200 billion parameters of GPT-4o).
Different LLMs, either more or less powerful, will likely produce different results.
Moreover, since GPT models are closed-source and version-controlled by OpenAI, we have no influence over potential future updates, which may lead to variations in output or performance.

Readability is assessed qualitatively by the authors, and while we aimed for objectivity, our judgment may not reflect that of other developers.
This is why we provide all generated test cases and our manual classification in the reproduction package.

\section{Future Work}

There are several promising directions for future research that build on the current findings and insights from related work.

One key avenue is to explore the use of open-source models, such as LLaMA 3.3, which have shown competitive performance to GPT-4o in recent evaluations.
Leveraging open-source models could enhance security by allowing for a fully on-premise deployment while also reducing costs.

Another important direction involves enhancing the feedback loop in the multi-agent system by integrating an automated version of the Restats tool.
By transforming Restats into an autonomous agent capable of analyzing test outcomes and dynamically guiding the Planner, the system could better target unexplored API behaviors and improve coverage.
This enhancement would allow the system to iteratively refine test strategies based on concrete coverage metrics, maximizing overall effectiveness.

Lastly, expanding the evaluation to cover real-world APIs in production environments and comparing the results with other approaches would provide valuable insights into the system’s robustness, scalability, and practical applicability.
Testing APIs with complex authentication flows, dynamic data dependencies, or real-time updates could reveal new challenges and guide further improvements to both agent strategies and system architecture.

\section{Conclusion}
This paper demonstrates how agentic LLM systems amplify REST API test cases.
We design and implement both single-agent and multi-agent systems.
Both systems achieve high structural API coverage and identification of many bugs in the system.
No human effort is required to design, write, and execute the test scenarios, except for a very simple happy-path test used as the initial test suite.
As expected, our multi-agent system outperforms our single-agent system in every category, though at higher computational, energy, and budget costs.

There are still opportunities to improve the outcomes. 
Prompting plays a crucial role; better-designed prompts may yield even better results.
The quality of OpenAPI documentation is another important aspect of the accuracy of the results. 
It is the primary knowledge base of our approach, and a more comprehensive OpenAPI documentation may yield better results.

In conclusion, while traditional automated REST API testing techniques provide valuable mechanisms for quality assurance, they often present notable limitations.
Methods like random-based or sample-value testing struggle to generate semantically valid inputs or understand complex dependencies between parameters.
The traditional tools often lack the capability for dynamic, iterative refinement based on execution feedback.
Our agentic LLM systems for REST API amplification address these gaps, increasing coverage with just a few targeted test cases, which enables a more qualitative, automated CI/CD pipeline.

These advancements can be a foundation for more intelligent and autonomous REST API software testing practices, supporting the continuous delivery and growing complexity of cloud-native applications.

\subsubsection{Acknowledgements}
This work is supported by the Research Foundation Flanders (FWO) via the BaseCamp Zero Project under Grant number S000323N.

%
%
%
%
\bibliographystyle{splncs04}
\bibliography{nooyens2025ictss}

\end{document}

%% file: usecommands.tex
\newcommand{\hypobox}[1]{
	\begin{center}%
        \noindent\thicklines\setlength{\fboxsep}{2pt}%
        \cornersize{0.1}
        \ovalbox{\begin{minipage}{12.0cm}%
		#1
		\end{minipage}}
	\end{center}}

\newcommand{\nb}[3]{}

\newcommand*{\RQTwo} [1] {And isn't this one even better ?}
\newcommand*{\RQThree} [1] {This one tops it all, doesn't it ?}

%% file: usepackages.tex
\usepackage{hyperref} 
\usepackage{amssymb} 
\usepackage{xspace}
\usepackage{color}
\usepackage{xcolor}
\usepackage{fancybox}
\usepackage{scrlayer-scrpage} 
\usepackage{ifthen} 
\usepackage{paralist} 

\usepackage[T1]{fontenc}
\usepackage{listings} 
\usepackage{comment}
\usepackage{array}
\usepackage{float}
\usepackage{adjustbox}
\usepackage{graphicx}
\usepackage{listings}
\usepackage{xcolor}
\usepackage{hyperref}
\usepackage{multirow}
\usepackage{hyperref}

\usepackage[normalem]{ulem} 

%% file: usesettings.tex
\definecolor{keywordcolor}{rgb}{0.0, 0.2, 0.6} 
\definecolor{commentcolor}{rgb}{0.3, 0.6, 0.3} 
\definecolor{stringcolor}{rgb}{0.2, 0.5, 0.2} 
\definecolor{classcolor}{rgb}{0.7, 0.5, 0.0} 
\definecolor{annotationcolor}{rgb}{0.6, 0.3, 0.0} 
\definecolor{numbercolor}{rgb}{0.4, 0.4, 0.4} 

\lstdefinestyle{javacode}{
  language=Java,
  basicstyle=\footnotesize\color{black}, 
  keywordstyle=\color{keywordcolor}, 
  commentstyle=\color{brown}\itshape, 
  stringstyle=\color{stringcolor}, 
  classoffset=1, 
  morekeywords={UploadImage}, 
  keywordstyle=[1]\color{classcolor}, 
  keywordstyle=[2]\color{annotationcolor}, 
  morekeywords=[2]{@Test}, 
  numbers=left,
  numberstyle=\tiny\color{numbercolor}, 
  stepnumber=1,
  frame=single,
  tabsize=4,
  showstringspaces=false,
  breaklines=true,
  captionpos=b,
  numbersep=10pt, 
  framexleftmargin=15pt, 
  xleftmargin=20pt 
}

\lstdefinelanguage{json}{
    basicstyle=\ttfamily\footnotesize\color{black}, 
    numbers=left, 
    numberstyle=\tiny\color{numbercolor}, 
    stepnumber=1, 
    numbersep=10pt, 
    showstringspaces=false, 
    breaklines=true, 
    frame=single, 
    tabsize=4, 
    framexleftmargin=15pt, 
    xleftmargin=20pt, 
    captionpos=b, 
    literate=
     *{0}{{{\color{blue}0}}}{1}%
      {1}{{{\color{blue}1}}}{1}%
      {2}{{{\color{blue}2}}}{1}%
      {3}{{{\color{blue}3}}}{1}%
      {4}{{{\color{blue}4}}}{1}%
      {5}{{{\color{blue}5}}}{1}%
      {6}{{{\color{blue}6}}}{1}%
      {7}{{{\color{blue}7}}}{1}%
      {8}{{{\color{blue}8}}}{1}%
      {9}{{{\color{blue}9}}}{1}%
      {:}{{{\bfseries\color{black}:}}}{1}%
      {,}{{{\bfseries\color{black},}}}{1}%
      {\{}{{{\color{black}\{}}}{1}%
      {\}}{{{\color{black}\}}}}{1}%
      {[}{{{\color{black}[}}}{1}%
      {]}{{{\color{black}]}}}{1}%
}

\hypersetup{
    colorlinks,%
    citecolor=black,%
    filecolor=black,%
    linkcolor=black,%
    urlcolor=gray,
    linktocpage=true,
    bookmarks=true,
    bookmarksopen=true,
    pdfpagemode=UseOutlines,
    pdftitle={Test Amplification for REST APIs via Single and Multi-Agent LLM Systems},    
    pdfauthor={Anonymous},     
    pdfsubject={Draft for ICTSS Conference},   
    pdfcreator={PDF LaTex},   
}